\documentclass[final,fleqn,10pt]{wlscirep}
\usepackage[utf8]{inputenc}
\usepackage[T1]{fontenc}
\title{A deep learning approach to identify unhealthy advertisements in street view images}

\usepackage{amssymb}
\usepackage{adjustbox}
\usepackage{pbox}
\usepackage{caption}
\usepackage{mathtools}
\usepackage{mwe}
\usepackage{tabularx}
\usepackage{enumitem}
\usepackage{soul}
\usepackage{algorithm}
\usepackage{algpseudocode}
\usepackage{pgfgantt}
\usepackage[utf8]{inputenc}
\usepackage[free-standing-units]{siunitx} 
\usepackage{bm}
\usepackage{balance}
\usepackage{listings}
\usepackage{multirow}
\usepackage{pdfpages}
\usepackage[graphicx]{realboxes}
\usepackage{balance}
\usepackage{setspace}
\usepackage[framemethod=tikz]{mdframed}
\usepackage[justification=centering]{caption}
\usepackage{subcaption}
\usepackage{graphicx}
\graphicspath{ {images/} }
\usepackage{afterpage}
\usepackage{float}
\usepackage{verbatimbox}
\newcommand{\ts}{\textsuperscript}
\definecolor{black}{rgb}		{0.0, 0.0, 0.0}
\definecolor{white}{rgb}		{1.0, 1.0, 1.0}
\definecolor{yellow}{rgb}		{1.0, 1.0, 0.8}
\definecolor{red}{rgb}			{0.6, 0.0, 0.2}
\definecolor{blue}{rgb}		{0.0, 0.2, 0.5}
\definecolor{green}{rgb}		{0.6, 0.8, 0.8}
\definecolor{dark_green}{RGB} {0, 140, 0}
\definecolor{gold}{rgb}		{0.6, 0.4, 0.1}
\definecolor{grey}{RGB}{0,0,0}
\definecolor{Gray}{gray}{0.8}
\definecolor{MediumGray}{gray}{0.9}
\definecolor{LightGray}{gray}{0.98}
\definecolor{LightCyan}{rgb}{0.88,1,1}
\definecolor{purple}{RGB}{128,0,128}
\definecolor{sl_blue}{RGB}{47, 60, 105}
\definecolor{orange}{RGB}{255,165,0}
\definecolor{Gray}{gray}{0.85}

\usepackage{xcolor,colortbl}

\newcolumntype{a}{>{\columncolor{grey}}c}
\newcolumntype{b}{>{\columncolor{white}}c}
\usepackage{lineno}

\author[1,4,+]{Gregory Palmer}
\author[1,*,+]{Mark Green}
\author[2]{Emma Boyland}
\author[3]{Yales Stefano Rios Vasconcelos}
\author[3]{Rahul Savani}
\author[1]{Alex Singleton}

\affil[1]{Geographic Data Science Lab, Department of Geography and Planning, University of Liverpool, UK}
\affil[2]{Department of Psychology, University of Liverpool, UK}
\affil[3]{Department of Computer Science, University of Liverpool, UK}
\affil[4]{L3S Research Center, Leibniz University Hannover, Germany}
\affil[*]{mgreen@liverpool.ac.uk}

\affil[+]{these authors contributed equally to this work}

\begin{abstract}
While outdoor advertisements are common features within towns and cities, they may reinforce social inequalities in health. Vulnerable populations in deprived areas may have greater exposure to fast food, gambling and alcohol advertisements encouraging their consumption. Understanding who is exposed and evaluating potential policy restrictions requires a substantial manual data collection effort. To address this problem we develop a deep learning workflow to automatically extract and classify unhealthy advertisements from street-level images. We introduce the Liverpool \ang{360} Street View (LIV360SV) dataset for evaluating our workflow. The dataset contains
25,349,
360 degree, street-level images collected via cycling with a GoPro Fusion camera, recorded Jan 14th -- 18th 2020. 10,106 advertisements were identified and classified as food (1335), alcohol (217), gambling (149) and other (8405) (e.g., cars and broadband). We find evidence of social inequalities with a larger proportion of food advertisements located within deprived areas and those frequented by students. 
Our project presents a novel implementation for the incidental classification of street view images for identifying unhealthy advertisements, providing a means through which to identify areas that can benefit from tougher advertisement restriction policies for tackling social inequalities. 
\end{abstract}
\begin{document}

\flushbottom
\maketitle
%
%
\thispagestyle{empty}



\section{Introduction}

The literature on advertising has previously shown that certain social demographics experience greater exposure to unhealthy products via a variety of advertisement platforms \cite{batada2008nine,isselmann2017sensitizing,powell2014racial,tatlow2017safe,adams2011socio}. There is increasing recognition of the role of unhealthy product consumption in the global non-communicable disease burden \cite{moodie2013profits}. In recent years, some public authorities have responded by introducing restrictions to limit exposure towards advertisements that encourage risky behaviour (e.g. Transport for London have banned all fast food advertisements on their networks). Understanding the populations and areas exposed to unhealthy advertisements, monitoring if regulations are being adhered to, and identifying areas to implement restrictions remain open problems. Collecting advertisement data within urban environments requires a substantial manual effort\cite{adams2011socio,liu2019space,kelly2008commercial,hillier2009clustering}\textcolor{black}{. As a result there are very few (if any) existing secondary datasets geolocating advertisements.} The rapid and dynamic nature of advertisements constantly changing also limits the use of surveying landscapes (which are time and cost intensive). 

The emergence of \emph{deep learning} \cite{Goodfellow-et-al-2016} for improved image classification raises the possibility of automating this task. 
Current state-of-the-art seamless segmentation networks \cite{Porzi_2019_CVPR} can be trained to identify billboards using the \emph{Mapillary Vistas Dataset} for semantic understanding of street scenes \cite{MVD2017}. However, this dataset does not account for different {content} categories. Furthermore, we consider that the manual annotation of advertisements within street-level imagery is both time consuming and can lead to a dataset with a limited shelf-life. 

Advertisement campaigns, company logos and product ranges are ever evolving \cite{gilbody2004direct}, rendering manual efforts obsolete. To mitigate this problem we present a workflow for extracting and classifying advertisements using an approach that is flexible and allows repeated data sweeps. 

The aim of our study is to develop a deep learning workflow to automatically extract and classify unhealthy advertisements from street view images. Our contributions can be summarized as follows:

\begin{enumerate}
\item We outline a novel and open workflow for extracting and classifying advertisements from street-level images. 
\item We introduce the open Liverpool \ang{360} Street View (LIV360SV) dataset, consisting of 25,349 geo-tagged street-level images for Liverpool, UK. Data will be updated longitudinally and the method can be deployed in varying contexts/environments.
\item We compare the clustering of extracted advertisements by socio-demographics to study the extent of social inequalities in unhealthy advertisement exposure.
\end{enumerate}


\section{Background} \label{sec:background}

\subsection{The Impact of Unhealthy Advertisements}

The Commercial Determinants of Health (CDoH), defined here as the processes where private organisations prioritise profit over public health, are powerful drivers of trends in non-communicable diseases and health inequalities \cite{kickbusch2016commercial,west2013commentary}.
Organisations may encourage the consumption of unhealthy products through marketing and advertisements campaigns across multiple platforms. There is a growing concern among public health officials regarding the number of advertisements for risky products e.g., alcohol, gambling, unhealthy food and beverages \cite{cassidy2017frequency,ireland2019commercial}. Numerous studies conducted around the world indicate that exposure to unhealthy energy‐dense, nutrition-poor food and beverage advertisements can promote unhealthy eating habits \cite{smits2015persuasiveness,lesser2013outdoor,calvert2020qualitative,sadeghirad2016influence,hershko2019advertising,martinez2018socioeconomic,walton2009examining}. The marketing of products that are high in fat, sugar and salt to children is particularly concerning, as it increases the potential for diet‐related diseases later in life \cite{sadeghirad2016influence}. Exposing adolescents to alcohol advertisements has been found to encourage early usage, and can lead to an increase in consumption \cite{anderson2009impact}, while gambling advertisements can trigger an impulse to increase activities, in particular in individuals who want to either quit or gamble less frequently \cite{binde2009exploring}.


\subsection{Differences in exposure to advertising}

When advertisements are prevalent within deprived areas, or areas with high levels of obesity, their role may counter public health efforts to tackle health inequalities. Evidence suggests a socio-economic difference in exposure to outdoor food advertising. For instance, in Newcastle upon Tyne, England larger spaces were found to be devoted to food advertisements within less affluent areas \cite{adams2011socio}. Differences in exposure meanwhile have been linked to a big data revolution, which has seen many firms possessing unprecedented amounts of information about consumers to enable advertisement campaigns to be aimed at individual demographics within the population \cite{johnson2013targeted,tatlow2017safe}. This practice has been shown to impact brand perceptions of the exposed demographic. Harris et al.\cite{harris2019qualitative} find that upon experiencing greater exposure towards advertisements promoting energy dense and nutrient poor foods, Black and Latino adolescents develop a more positive attitude towards the promoted brand. Pasch et al. \cite{pasch2009does} show that the number of outdoor alcohol advertisements found within 1500 feet of 63 Chicago schools is significantly higher for schools with 20\% or more Hispanic students -- 6.5 times higher than for Schools with less than 20\% Hispanic students. Alcohol marketing campaigns have also been shown to be more prevalent around areas frequented by University students. Kuo et al.\cite{kuo2003marketing} find that alcohol advertisements are prevalent in the alcohol outlets around college campuses in the USA.



Students are also a demographic more likely to be exposed to gambling advertisements. Clemens et al. \cite{clemens2017exposure} find that high exposure towards gambling advertisements 
is positively related to all assessed gambling outcomes. In addition, strong associations have been found for adolescents and students engaging in risky behaviour such as drinking and gambling when exposed to related advertisements \cite{jones2011exposure,lopez2018alcohol}. Problem gambling in particular has the potential to be amplified by drinking and eating disorders. Lopez et al. \cite{lopez2018alcohol} investigate the extent to which gambling commercials are promoting risky behaviour of drinking alcohol and eating low nutritional value food, looking at the narratives depicted within the advertisements. The authors find that British and Spanish football betting advertisements attempt to align the consumption of alcohol with sports culture and friendship within the emotionally charged context of watching sporting events. Indeed, even far reaching sporting bodies, e.g., the English Premier League, have been shown to have marketing portfolios that include unhealthy products \cite{ireland2019commercial}.


Restricting exposure to unhealthy advertisements meanwhile has been found to have a positive effect on behaviour \cite{walton2009examining}. Lwin et al. \cite{lwin2020macro}, for example, study the impact of food advertising restrictions enforced in Singapore. The authors find that children's cognition towards fast-food shifts in a desirable direction upon a stricter policy being adopted, with household stocks of unhealthy food items also decreasing. However, while there is evidence that vulnerable populations are more exposed to unhealthy advertisements and restricting them is an effective strategy, much of these data come from lab-based studies. 

{To our knowledge, there are very few to no known data available on the location of outdoor advertisements. While advertisements range in type (e.g. online, print, vehicles), outdoor advertisements are prominent features of environments that individuals may experience and interact with in their everyday experiences. The lack of available data represents a significant gap in our ability} to be able to understand differential patterns of exposure, as well as effectively evaluative the impact of future regulative interventions. We need effective and efficient data systems that map advertisement locations. Traditional data collection strategies employ primary surveys to locate advertisements, however such methods are time and cost intensive making them static snapshots that fail to capture the dynamic and evolving aspects of advertisement strategies. 

\subsection{Deep Learning}

Utilising incidental data sources, coupled with maturing image classification techniques\textcolor{black}{,} offers one way forward to improve and automate the data collection process efficiently. Deep Learning is one technique that has shown a lot of promise for developing solutions to challenging virtual and real world problems \cite{silver2017mastering,mnih2013playing}. These successes can be attributed to breakthroughs that enable \emph{deep neural networks} to learn solutions to problems that humans solve using intuition \cite{Goodfellow-et-al-2016}. Deep neural networks are trained to extract compact features from complex high dimensional input data. They accomplish this by combining layers of hierarchical features into ever more complex concepts. Our workflow uses Convolutional Neural Networks (ConvNets), which can extract features from inputs in the form of arrays and tensors \cite{lecun2015deep}. A ConvNet trained to classify images consists of layers of neurons, with the first layer extracting edges, which are combined into corners and contours by the next layers, before subsequently being combined to form the object parts that enable a classification. Through stacking multiple non-linear layers the network can be trained using stochastic gradient descent to implement complex functions, that are sensitive towards minute details within inputs, while simultaneously being able to ignore less relevant features \cite{lecun2015deep}. Through building an effective classifier that can be updated with new information (important when advertisements are constantly changing), deep learning offers a deployable tool that automatically classify images more efficiently than manual coding by researchers. 

\section{Data} \label{sec:data}

\subsection{Mapillary Vistas dataset}


Street level images (also known as street view images) are panoramic images recorded at set intervals. Services such as Google Street View, Bing Maps and Mapillary use these data to provide a virtual representation of map locations. \textcolor{black}{In 2017 Mapillary introduced the Vistas dataset \cite{MVD2017} to aid the development of state-of-the-art methods for road scene understanding. The dataset consists of 25,000 densely-annotated, internationally crowd sourced, street level-images with 66 object categories, including \emph{billboards}. Approximately 90\% of the images are from road / sidewalk views in urban areas, with the remaining being rural areas and off-road.} Individual objects within each images are delineated using polygons. Since its release the Mapillary Vistas has frequently been used for benchmarking panoptic street scene segmentation methods \cite{Kirillov_2019_CVPR,Porzi_2019_CVPR}. 

\subsection{The Liverpool 360 Street View Dataset}

\begin{figure}[h]
\centering
\includegraphics[width=0.73\columnwidth]{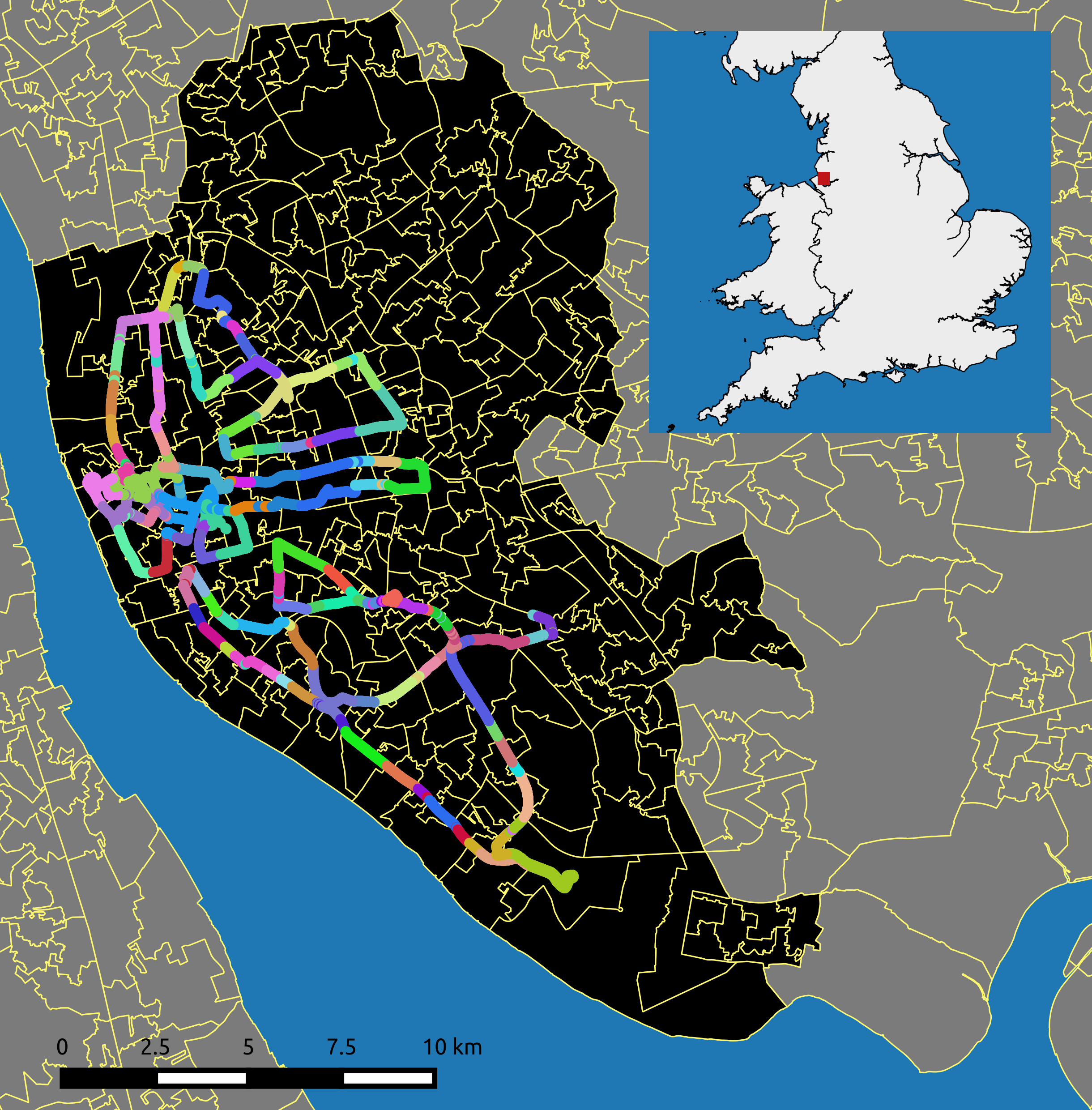}
\caption{Map depicting the coverage of the LIV360SV dataset color coded by lower-layer super output areas (LSOAs). \\ This map was created using QGIS 2.8.6-Wien~\cite{qgis}.}
\label{fig:LIV360SV}
\end{figure}

While there exists an abundance of street-level imagery on platforms such as \emph{Google Street View}, the recently imposed costs for using Google's API, as well as cases of Google updating terms and conditions to hinder researchers, highlights the need for alternative open sourced solutions. 
Existing open and crowd sourced street-level images predominately lack the quality of the interactive panoramas found on services such as Google Street View. Images are frequently recorded using dashboard cameras, and as a result have a restricted field of vision. Motivated by these factors we record an open street-level dataset for Liverpool, using a GoPro Fusion \ang{360} camera attached to a member of the team (Mark Green) who cycled along major roads. We follow Mapillary's recommendations for recording street-level images (\href{https://help.mapillary.com/hc/en-us/articles/360026122412-GoPro-Fusion-360}{https://help.mapillary.com/hc/en-us/articles/360026122412-GoPro-Fusion-360}). The camera records front and back images at 0.5 second interval, which we later stitch together using GoPro Fusion Studio. To date our dataset consists of 25,349 street-level images each with GPS location recorded. We illustrate the current coverage of the LIV360SV dataset in Figure~\ref{fig:LIV360SV}. 
We focused on sampling three areas of Liverpool with varying contexts over three different days: (1) City Centre (Jan 14\ts{th} 2020) - areas characterised by shops and services; (2) North Liverpool (Jan 15\ts{th} 2020) - areas contain high levels of deprivation; (3) South Liverpool (Jan 18\ts{th} 2020) - areas include a mixture of affluent populations and diverse ethnic groups (See \href{https://www.mapillary.com/app/org/gdsl_uol?lat=53.39&lng=-2.9&z=11.72&tab=uploads}{https://www.mapillary.com/app/org/gdsl\_uol?lat$=$53.39\&lng$=$-2.9\&z$=$11.72\&tab$=$uploads}). 

\subsection{Advertisement Data} \label{sec:google_images_ds}


The Mapillary Vistas includes a billboards category {that we utilise here. In practice, the seamless segmentation network's billboard category extracts billboards, company logos (e.g., the McDonald's arches) and store front signs. We therefore consider each of these items to be advertisements within the context of this study, capturing the potential range of unhealthy imagery individuals may be exposed to and influenced by.}
Furthermore, the dataset does not distinguish different types of {content (e.g. advertisement type)}. Further annotations would therefore be necessary to train panoptic scene segmentation networks to differentiate between advertisement types. 
Manually annotating segmentation masks is a time consuming task. {Instead, we propose to classify advertisements extracted from street level images using a model trained to classify advertisement images. To train the classifier we manually label advertisements extracted from a neighbouring city, Manchester, UK, which we also download from Mapillary.
While these data typically originate from dashboard cameras, and are therefore likely to miss advertisements within an environment, the extracted advertisements are adequate to train a classifier to distinguish content categories.
Manchester was selected since it is geographically close to Liverpool, as well having a similar historical context (i.e. Northern industrial city with high levels of deprivation spatially concentrated) that may see similar types of advertisements.}


\subsection{Spatial data} \label{sec:spatial_data}

To examine the extent of geographical clustering in the socio-demographic types of areas that advertisements are located, we use two area level datasets. 

\emph{First}, neighbourhood deprivation is measured using the English Indices of Deprivation 2019 \cite{ucl_ioe34258}. The index measures neighbourhood deprivation based on seven domains including income, employment, education, health, crime, access to housing and services, and environmental features. Data are measured for Lower Super Output Areas (LSOAs) which are administrative zones with an average population size of $\approx1500$ people. We use decile of deprivation rank for analyses.

\emph{Second}, socio-demographic area type is measured using 2011 Output Area Classification (OAC)~\cite{gale2016creating}. OAC is a neighbourhood classification built using data from demographic (e.g. age, sex, ethnicity) and social (e.g. occupation, education) measures to classify `area types'. OAC comprises 8 Supergroups and 26 Groups which we describe in Table~\ref{tab:OAC:cluster_names}. We focus our evaluation at the Supergroup and Group levels. Output Areas are administrative zones with a minimum of 100 people.

\begin{table}[h]
\centering
\resizebox{0.48\columnwidth}{!}{%
\begin{tabular}{ | l | l |}
\hline
\textbf{Super-Group} & \textbf{Group} \\
\hline
\multirow{ 3}{*}{\textbf{1} - Rural Residents} & \textbf{1a} - Farming Communities \\
& \textbf{1b} - Rural Tenants \\
& \textbf{1c} - Ageing Rural Dwellers \\
\hline
\multirow{ 4}{*}{\textbf{2} - Cosmopolitans} & \textbf{2a} - Students Around Campus \\
& \textbf{2b} - Inner-City Students  \\
& \textbf{2c} - Comfortable Cosmopolitans \\
& \textbf{2d} - Aspiring and Affluent \\
\hline
\multirow{ 4}{*}{\textbf{3} - Ethnicity Central} & \textbf{3a} - Ethnic Family Life \\
& \textbf{3b} - Endeavouring Ethnic Mix \\
& \textbf{3c} - Ethnic Dynamics \\
& \textbf{3d} - Aspirational Techies \\
\hline
\multirow{ 3}{*}{\textbf{4} - Multicultural Metropolitans} & \textbf{4a} - Rented Family Living \\
& \textbf{4b} - Challenged Asian Terraces \\
& \textbf{4c} - Asian Traits \\
\hline
\multirow{ 2}{*}{\textbf{5} - Urbanites} & \textbf{5a} - Urban Professionals and Families \\
& \textbf{5b} - Ageing Urban Living \\
\hline
\multirow{ 2}{*}{\textbf{6} - Suburbanites} & \textbf{6a} - Suburban Achievers \\
& \textbf{6b} - Semi--Detached Suburbia \\
\hline
\multirow{ 4}{*}{\textbf{7} - Constrained City Dwellers} & \textbf{7a} - Challenged Diversity \\
& \textbf{7b} - Constrained Flat Dwellers \\
& \textbf{7c} - White Communities \\
& \textbf{7d} - Ageing City Dwellers \\
\hline
\multirow{ 4}{*}{\textbf{8} - Hard--Pressed Living} & \textbf{8a} - Industrious Communities  \\
& \textbf{8b} - Challenged Terraced Workers \\
& \textbf{8c} - Hard--Pressed Ageing Workers\\
& \textbf{8d} - Migration and Churn \\
\hline
\end{tabular}} 
\caption{Area classification for output area (OAC) cluster names \cite{gale2016creating}.}
\label{tab:OAC:cluster_names}
\end{table}


\section{Method} \label{sec:method}

\begin{figure}[h]
\centering
\includegraphics[width=0.85\columnwidth]{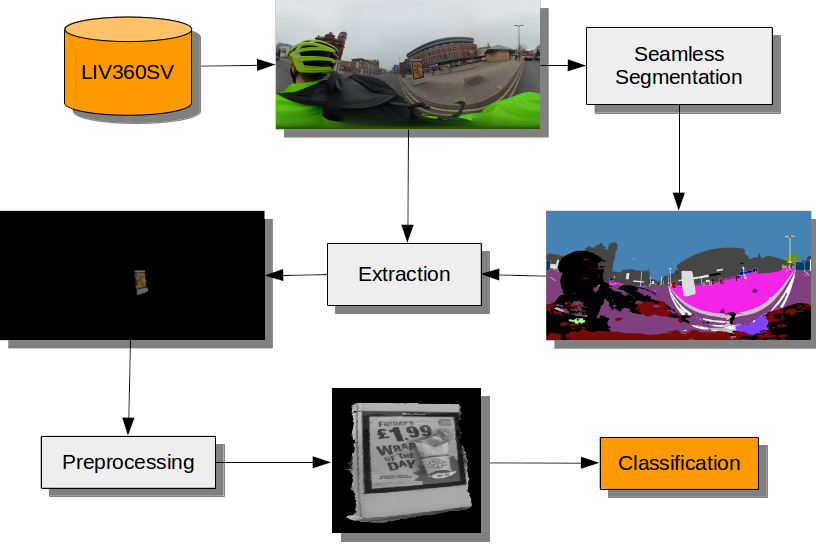}
\caption{An illustration of our advertisement classification workflow. (The final image was converted to monochrome for the Figure – colour images are used in modelling)}
\label{fig:Architecture_Diagram}
\end{figure}

Figure~\ref{fig:Architecture_Diagram} illustrates our workflow, and we discuss each individual component in detail below. For implementation details and dataset download instructions visit: \href{https://github.com/gjp1203/LIV360SV}{https://github.com/gjp1203/LIV360SV}.

\subsection{Seamless Scene Segmentation}



For extracting advertisements from street level images we use the seamless scene segmentation network introduced by Porzi et al. \cite{Porzi_2019_CVPR}. The network offers advantages of both semantic segmentation -- determining the semantic category that a pixel belongs to -- and instance-specific semantic segmentation -- the individual object that a pixel belongs to, enabling differentiation between neighbouring entities of the same type. The authors achieve state-of-the-art results on three street-view datasets, including Cityscapes \cite{Cordts2016Cityscapes}, the Indian Driving Dataset \cite{varma2019idd} and Mapillary Vistas \cite{MVD2017}. \textcolor{black}{We use the implementation provided by Porzi et al. \cite{Porzi_2019_CVPR} -- which has been pre-trained on the Mapillary Vistas dataset -- to assign semantic categories to our images. Subsequently areas that have been categorised by the network as type 'billboard' are extracted from the images.} We find that the pre-trained network achieves a mean intersection over union (mIoU) score of 0.397 for the category billboard on the Mappilary Vistas validation set. However, larger mIoU scores are achieved when setting a threshold for the minimum number of billboard pixels for considering an image, reaching similar values to the 0.458 achieved in Porzi et al. \cite{Porzi_2019_CVPR} across categories~(see supplementary material). In addition we evaluate the network's error with regards to falsely detected and missed ads. The ground truth labelling of the Mapillary Vistas validation set contains 4,562 billboards of 2,000 pixels and above -- our selection criteria for extracting advertisements, as for smaller advertisements it is often challenging to assign a category label. In comparison the semantic scene segmentation network extracted 4,305 billboards, as well as 143 items that were falsely classified as billboards, consisting of street signs, blank surfaces, traffic lights, and interestingly clock faces. 

\subsection{Extraction}


{Upon using the seamless scene segmentation network to identify regions within a street level image that have been categorised as type billboard, we first use OpenCV's  \cite{itseez2015opencv} 'connected components with stats method' to identify individual advertisements within a street level image (as each street level image may contain multiple advertisements). We draw a convex hull around each disjoined group of billboard pixels (i.e., each individual advertisement identified within the image) and fill the polygon to obtain a binary mask. Polygons containing fewer than 2,000 pixels are discarded, as the majority of advertisements of this size are difficult to assign a label to. If an image contains multiple advertisements we obtain a binary mask for each advertisement with their respective locations. Within these individual masks, ones indicate areas within the convex hull drawn around the individual pixels identified as type billboard, while zeros mask the remaining entities within the images. The masks subsequently allow us to extract the individual advertisements one at a time.}

\subsection{Preprocessing} \label{sec:preprocessing}

{We \textcolor{black}{divide} pre-processing into two steps. First we perform a number of operations to crop and spatially transform the images to a frontal view. We subsequently outline an approach towards dealing with over-representation, resulting from the same advertisement being extracted multiple times from temporally sequential images.}

{\textbf{Obtaining a frontal view:}} With the remaining content having been masked out during the extraction step we subsequently crop the images. {However, we observe that distinct billboards depicting the same advertisement will often be recorded from a different point of view, for instance due to differences in the location of each billboard and road layouts. Therefore, as the final step of our workflow is to pass the extracted items to a classifier, we take an addition step of training a Spatial Transformation Network (STN) \cite{jaderberg2015spatial} to transform the extracted items to a frontal view, thereby increasing the likelihood of training and testing images having a similar depiction.}  

{\textbf{Addressing the duplication of distinct advertisements:} Recording street level images at 0.5 second intervals brings the risk of recording distinct advertisements multiple times. As a result natural obstacles within the environment (e.g., traffic) can lead to some advertisements enjoying a greater representation compared to others. Implementing spatial constraints meanwhile -- such as only evaluating images taken every $n$ meters -- can result in less visible advertisements being missed. To address this issue we propose a method towards identifying the duplication of distinct advertisements. Our approach involves measuring the similarity of advertisements extracted from spatially proximate street level images within a specified Euclidean distance $d$. We use Scale-Invariant Feature Transform (SIFT) \cite{lowe1999object} to detect matching features within pairs of advertisements. We subsequently construct a graph $G$ where the nodes represent the extracted advertisements. Edges are added between nodes where the number of matching features exceeds a threshold $\tau$. We treat each disconnected sub-graph $g \in G$ as a distinct advertisement. Where $|g| > 1$ we discard all but the advertisement located closest to the centroid of $g$.}

{Considerations are required regarding the setting of the distance limit $d$ and matching features threshold $\tau$. While calibrating our method we found evidence that low values for $\tau$ and larger values for $d$ result in two distinct advertisements being assigned to the same sub-graph $g$. However, using large values for $\tau$ reduces the number of true positives. Based on the experiments outlined in the supplementary material we conduct our evaluation below using $\tau = 60$ and $d=10m$. While some duplication remains with this setting, it allows us to automatically identify the largest instances of duplication, resolving the imbalance within the data used for our evaluation.}

\subsection{Classification} \label{sec:method:classifier}


{We classify extracted advertisements using Keras' InceptionV3~\cite{szegedy2016inception} implementation with weights pretrained on imagenet. We train the network for five 100 step epochs, using a learning rate of 1e-4 and a batch size of 32 images per step. The inputs images are of size 224x224 pixels. We also apply a common dataset augmentation technique of adding random rotations (with a 30 degrees limit) when sampling images. We accelerate the training process using a GeForce GTX 1080 GPU.}

\section{Results} \label{sec:results}

We take a two-step approach towards evaluating our proposed workflow. First we analyse the clustering of advertisements extracted using the seamless scene segmentation network component. For precision we conduct this analysis upon assigning ground truth labels to the extracted advertisements. Our second step is to evaluate the extent to which an {InceptionV3} network can be trained to classify the extracted advertisements. 

\subsection{Examining inequalities in advertisement locations} \label{sec:results:spatial}

\begin{figure}[h]
\centering
\includegraphics[width=0.78\columnwidth]{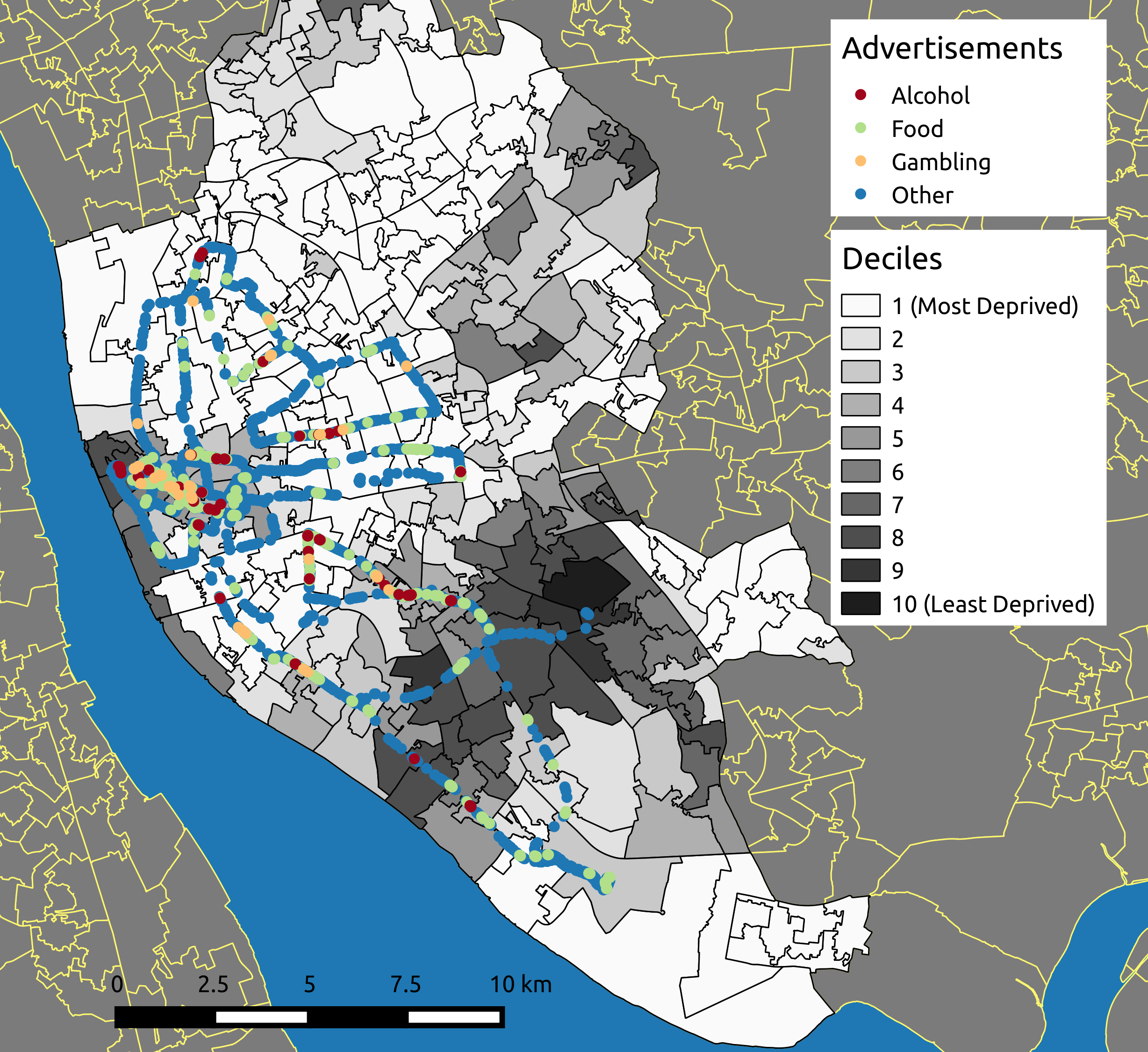}
\caption{Liverpool advertisement locations by Lower Super Output Areas (LSOAs). A color gradient indicates the level of deprivation, with white and black being the most and least deprived respectively. This map was created using QGIS 2.8.6-Wien~\cite{qgis}.}
\label{fig:Deprivation}
\end{figure}

{We identified 10,106 advertisements, classified as \emph{food} (1335), \emph{alcohol} (217), \emph{gambling} (149) and \emph{other} (8405).
Upon removing near duplicate advertisements using the method outlined in Section \ref{sec:preprocessing} we are left with \emph{food} (873), \emph{alcohol} (102), \emph{gambling} (79) and \emph{other} (6247) advertisements.} In Figure~\ref{fig:Deprivation} we illustrate the distribution of advertisements belonging to each category across the LSOAs for Liverpool. The LSOAs are each assigned a color shading based on the decile that they belong to, with white and black representing the most and least deprived respectively. Advertisements are represented by circles. 

\begin{figure}[h]
\centering
\begin{subfigure}[b]{0.32\columnwidth}
    \centering
    \includegraphics[width=\columnwidth]{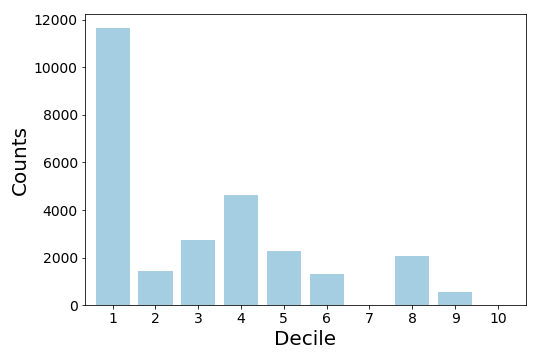}
    \caption{Street-level image totals per deprivation decile.}
    \label{sub-fig:Deprivation:dep_decile_counts}
\end{subfigure}
\begin{subfigure}[b]{0.32\columnwidth}
    \centering
    \includegraphics[width=\columnwidth]{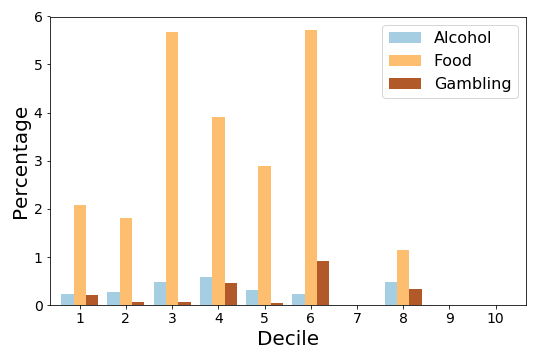}
    \caption{Unhealthy advertisement percentages.}
    \label{sub-fig:Deprivation:dep_unhealthy_gt_occurences}
\end{subfigure}
\begin{subfigure}[b]{0.32\columnwidth}
    \centering
    \includegraphics[width=\columnwidth]{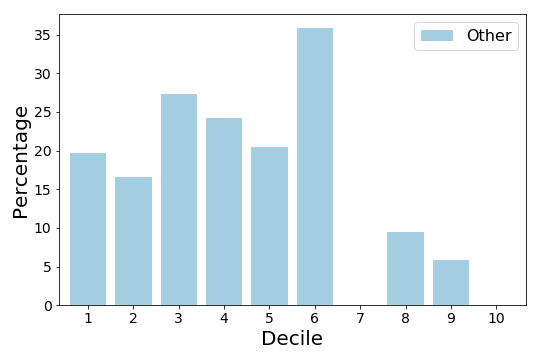}
    \caption{Other advertisement percentage.}
    \label{sub-fig:Deprivation:dep_other_gt_occurences}
\end{subfigure}    
\caption{Sub-Figure \ref{sub-fig:Deprivation:dep_decile_counts} illustrates the number of street-level images per deprivation decile according to the 2019 English indices of deprivation. Sub-Figures \ref{sub-fig:Deprivation:dep_unhealthy_gt_occurences} and \ref{sub-fig:Deprivation:dep_other_gt_occurences} contain the percentage of images with unhealthy advertisements and those of type `other' respectively.}
\label{fig:Deprivation:bar_plots}
\end{figure}

We turn to bar-plots in Figure~\ref{fig:Deprivation:bar_plots} to illustrate exposure towards unhealthy advertisements per decile of deprivation. However, in Sub-Figure~\ref{sub-fig:Deprivation:dep_decile_counts} we observe an imbalance in the number of street-level image samples per decile within the LIV360SV dataset. We therefore focus on the proportion of advertisements found within each decile. In Sub-Figure~\ref{sub-fig:Deprivation:dep_unhealthy_gt_occurences} we observe that, the less deprived LSOAs have \textcolor{black}{proportionally} fewer advertisements compared to the more deprived areas. Larger proportions of food advertisements are found within deciles 1 to 6. For gambling meanwhile the largest proportion of advertisements are found within decile 6. {We tested whether the differences we observed across deprivation decile were meaningful (Table 2). While we found statistically significant differences across deciles for alcohol, food and other advertisements supporting our observation that they were more common in deprived areas, we found no association for gambling.}

\begin{figure}[h]
    \centering
    \begin{subfigure}[b]{0.32\columnwidth}
    \centering
    \includegraphics[width=\columnwidth]{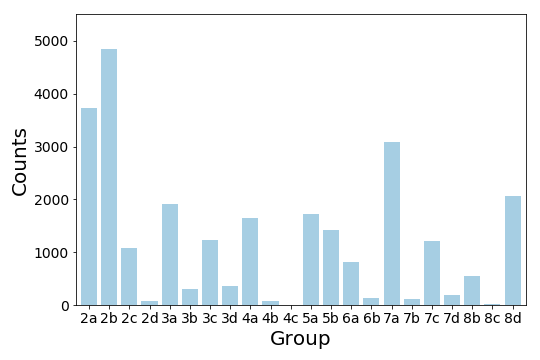}
    \caption{Image totals per OAC category.}
    \label{sub-fig:OAC:oac_totals}
    \end{subfigure}
    \begin{subfigure}[b]{0.32\columnwidth}
    \centering
    \includegraphics[width=\columnwidth]{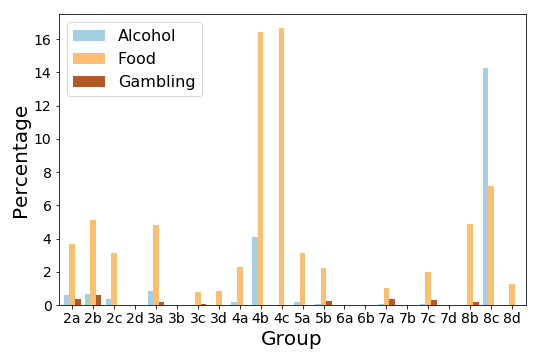}
    \caption{Unhealthy advertisement percentages.}
    \label{sub-fig:OAC:oac_GRP_unhealthy_gt_occurences}
    \end{subfigure}
    \begin{subfigure}[b]{0.32\columnwidth}
    \centering
    \includegraphics[width=\columnwidth]{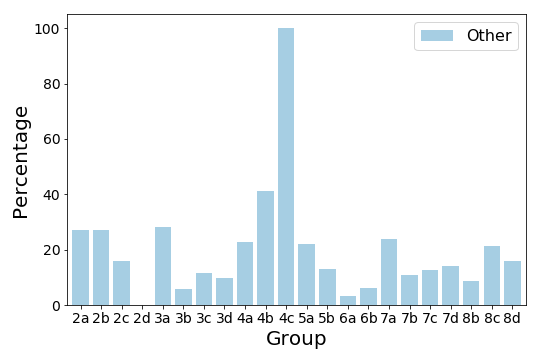}
    \caption{Other advertisement percentage.}
    \label{sub-fig:OAC:oac_GRP_other_gt_occurences}
    \end{subfigure}    
    
\caption{Sub-Figure \ref{sub-fig:OAC:oac_totals} illustrates the number of street-level images per OAC category. Sub-Figures \ref{sub-fig:OAC:oac_GRP_unhealthy_gt_occurences} and \ref{sub-fig:OAC:oac_GRP_other_gt_occurences} contain the percentage of images per OAC that contain either unhealthy advertisements and those of type `other' respectively.}
\label{fig:oac}
\end{figure}

Figure~\ref{fig:oac} compares the proportions of advertisements by OAC area type. For alcohol we observe that a large proportion of advertisements belong to OAC 8c -- Hard Pressed Aging Workers (14.29\%, see Sub-Figure~\ref{sub-fig:OAC:oac_GRP_unhealthy_gt_occurences}). However, this category only contains 14 images (Sub-Figure \ref{sub-fig:OAC:oac_totals}). Among the better represented categories the largest proportions of advertisements can be found within 2a -- Students Around Campus ({0.59\%}), 2b -- Inner City Students ({0.70\%}), 3a -- Ethnic Family Life ({0.83\%}) and 4b -- Challenged Asian Terraces ({4.11\%}). For gambling large proportions of advertisements are also located within 2a ({0.35\%}) and 2b ({0.64\%}). We also observe larger proportional representation under Super-Group 7 -- Constrained City Dwellers, in particular 7a -- Challenged Diversity (0.36\%) and 7c -- White Communities ({0.33\%}). The largest proportions of food advertisements can be found within super-groups 2 - Cosmopolitans, 4 - Multicultural Metropolitans and 8 - Hard-Pressed Living. Specifically, 2a - Students Around Campus ({3.67\%}), 2b- Inner-City Students ({5.11\%}), 2c-Comfortable Cosmopolitan ({3.12\%}), 4b-Challenged Asian Terraces ({16.44\%}), 4c-Asian Traits (16.67\%), 8b-Challenged Terraced Workers ({4.85\%}) and 8c-Hard–Pressed Ageing Workers (7.14\%). However, 4b (73), 4c (6) and 8c (14) contain less images compared to the other categories. {These differences observed across OAC groups were statistically significant for all advertisement categories (Table 2).}

\begin{table}[H]
\centering
\begin{tabular}{ | c | c | c |}
\hline
\textbf{Advert} & \textbf{Deprivation} & \textbf{OAC} \\
\hline
Alcohol & 70.99*** & 120.62*** \\
\hline
Food & 686.16*** & 301.97*** \\
\hline
Gambling & 11.16 & 25.14*** \\
\hline
Other & 628.95*** & 976.67*** \\
\hline
\end{tabular}
\label{tab:pvalues}    
\caption{Chi-squared tests for patterns in advert location by deprivation decile and output area classification (OAC) group. Note: p-values are denoted as *<0.05, **<0.01, ***<0.001}
\end{table}

\subsection{Towards Automated Classifications}~\label{sec:results:automated_classifications}

{The above examination of inequalities in advertisement locations was conducted with manually labelled samples. However, to conduct our evaluation at scale (e.g., for the entire UK) manually labelling samples would represent an obvious time consuming bottleneck in our framework. Therefore, we now evaluate if advertisements extracted from street-level imagery can be categorised automatically using a state-of-the-art image classifier. As outline in Section~\ref{sec:method:classifier} we train an Inception-V3 network using a training dataset that consists of manually labelled advertisements extracted from Mapillary street view images for the city of Manchester. We obtain these advertisements using the seamless segmentation network from Porzei et al. \cite{Porzi_2019_CVPR}, essentially repeating the process that we conducted for Liverpool. Via this process we obtained 3,096 food, 197 alcohol, 141 gambling and 34,198 other images. \textcolor{black}{We note however that due the majority of these images having been extracted from dash-cam footage, the quality was lower compared to those extracted from our LIV360SV data-set.}} 

{Our hypothesis is that the advertisements found within a neighboring cities are similar, giving us a source training dataset that is closely aligned with our target dataset, i.e., the advertisements extracted from LIV360SV. However, the Manchester dataset also has a shortage of images for the categories alcohol and gambling. We therefore focus our evaluation on the categories food and other, leaving the alcohol and gambling categories for future work. Considerations regarding how these imbalances could be addressed are provided in future studies within the discussion section below. We use oversampling to address the imbalance between the categories food and other.}

{Upon training the InceptionV3 network using the Manchester dataset we obtain weighted precision, recall and F1 scores of 
0.8518, 0.7157 and 0.7608 respectively. We use the weighted version of Keras' precision recall F1 score support function to address the imbalance in the number of samples for food and other images extracted from LIV360SV. In Table \ref{tab:pandr} we provide category wise mean precision, recall and F1 scores after randomly assigning samples from other to five subsets of size equal to food. We observe high precision for food, with fewer advertisements from category other being classified as food. However, food images are often classified as category other, explaining the lower recall score. We hypothesize that these scores can be further improved when training a classifier with a large-scale high-quality variation of our current training-set, for instance through extracting images from additional cities within the region. For a qualitative evaluation of the input features determining the classifications we refer the reader to Section 4 within the supplementary material.}

\begin{table}[h]
\centering
\begin{tabular}{ | c | c | c | c |}
\hline
\textbf{Category} & \textbf{Precision} & \textbf{Recall} & \textbf{F1} \\
\hline
Food & 0.76 & 0.619 & 0.68 \\
\hline
Other & 0.662 & 0.787 & 0.718 \\
\hline
\end{tabular}
\caption{Precision, recall and F1 scores for the automated classification component of our framework.}
\label{tab:pandr}    
\end{table}

\section{Discussion}


Our study demonstrates a novel workflow that can be used to efficiently identify the location of unhealthy advertisements from street-view imagery. To date we have extracted 10,106 advertisements for Liverpool, UK, categorised as food (1335), alcohol (217), gambling (149) and other (8405). There was distinct geographical clustering of advertisements particularly with greater amounts of unhealthy advertisements in deprived areas and student populations. Our approach addresses the dearth of data available on the location of unhealthy advertisements, offering an efficient and deployable tool for surveying other towns and cities.

The prevalence of food, gambling and alcohol advertisements within areas classified as inner-city students and campus provides further evidence that the student population is experiencing greater exposure to advertisements for unhealthy products\cite{jones2011exposure,lopez2018alcohol}.
 Regulating these areas and protecting younger student populations might be a key policy goal particularly as this period of the life course is important at establishing behaviours that may continue into later life. The clustering of unhealthy food advertisements in deprived areas may exacerbate inequalities in obesity and related health conditions. This would suggest that any policy to regulate the location of unhealthy food advertisements would be progressive and potentially help to narrow inequalities.


Having identified the prevalence of unhealthy advertisements within areas frequented by students opens up interesting avenues for future research. For example, given advertisers' attempts to normalize the consumption of unhealthy items while gambling with friends \cite{lopez2018alcohol}, an evaluation could be conducted to determine whether these behaviours are more likely to be enacted in areas with greater exposure. In addition, insights could be gained through differentiating between advertisement formats and studying the extent to which each type contributes towards triggering an impulse to gamble, e.g., billboard, shop window, and store signs.

{A further avenue for future research is to evaluate how exposures to unhealthy advertising vary in relation to policy interventions and strategies. For example, evaluating} the extent to which the current rules restricting the promotion of high fat, sugar and salt (HFSS) products within 100 meters from schools is deterring advertisers  (\href{https://www.asa.org.uk/advice-online/food-hfss-media-placement.html}{https://www.asa.org.uk/advice-online/food-hfss-media-placement.html}). In addition, we consider that individuals are often exposed to advertisements via dynamic entities. Bus stops for instance use monitors that can switch between advertisements. Developing our approach to account for these issues will be useful for future research. Further, insights could be gained through differentiating between advertisement formats and studying the extent to which each type contributes towards triggering behaviours to identify where regulations should focus their efforts.



A key strength to our study is the efficient data collection of advertisement locations. {We make methodological advancements in measuring the location of unhealthy advertisements through utilising a novel deep learning approach, with no known prior research applying similar methods or developing efficient alternatives.} To our knowledge, there is no open dataset that charts the location of advertisements in the UK. Having access to open data on advertisement locations is key for making effective policy decisions. Through automating the classification of street-view imagery, our approach can be efficiently combined with incidental data sources to locate advertisements over time with little additional time or resource costs. Expanding our data collection efforts to additional cities will help improve data coverage. {This could be supplemented through crowd sourcing images through recruiting and allowing individuals to take photos and geo-tag advertisements using an app. Increasing the number and balance of advertisement types could improve our model performance especially for those advertisements with low counts (alcohol and gambling).}



There are several limitations with regards to both the data and methods used in this paper. First, LIV360SV contains a number of unhealthy advertisements that are worthy of their own category. For instance, electronic cigarettes and vaping devices have become the most common tobacco products used by youth, with brands using similar marketing and advertising strategies as previously used for traditional tobacco products~\cite{walley2019public}. Classifying new categories would require retraining our classifier using additional data. Similarly, when applying our approach to a different location representative training data must be obtained for local brands and product ranges. Although our `other' category may not be specific, it captures the total potential exposure for unhealthy advertisements given that advertisements may change weekly in their content.

We note that the data collection process requires a systematic approach. Figure \ref{fig:Deprivation:bar_plots} displays that our dataset is skewed towards more deprived areas with regards to the number of samples. {This reflects both the historical concentration of deprivation in Liverpool, as well the lack of a systematic routing approach to data collection. Cycling routes should be designed to ensure representative routes (e.g. using local demographic data alongside GIS network routing methods).} Collecting data across different contexts and cities will also help to improve the generalisability of our dataset. Our initial data collection wave was in January where anecdotally during data collection, many advertisements were observed as relating to gyms or physical exercise. Commercial firms may release advertisements at different parts of the year based on seasonal trends (e.g. Easter and chocolate), events (e.g. gambling around sporting events) or product development. We plan to record seasonal data to enable a longitudinal study of advertisements within Liverpool. 


{While deep learning enables new possibilities with regards to evaluating our environments, we note that any conclusions drawn from evaluations underpinned by this technology should be cautious. As can be seen within our data-set, misclassifications exist with regards to false-positives/negatives, incorrect classification in mixed imagery (e.g. advertisements containing both fast food meals and alcohol) or advertisements missed from the workflow. We argue that our work shows the potential of these technologies in generating valuable data on exposures of unhealthy advertisements where such data does not exist. Future methodological work should seek to refine these approaches to improve their utility for informing public health initiatives.}

{We consider how exposure to advertisements relates to static populations (e.g. deprivation), however populations move around cities meaning that residential neighbourhoods are not always the best measure for experiences. Extending our analyses to assess how exposure to unhealthy advertisements varies by population flows (e.g. commuting patterns along major roads, differences in day- and night-time populations). Advertisements may be targeted at these flows to maximise their potential audiences. Understanding these inequalities in dynamic exposures may identify particular spaces that are important for elucidating exposures (e.g. work, school), which could be targeted by interventions.} 



Finally, steps are necessary to improve the accuracy of the workflow's classifier component (Section~\ref{sec:results:automated_classifications}). Our evaluation shows that our approach requires more representative training images for food, and as mentioned, we only managed to collect a limited number of advertisements of type gambling and alcohol. {We note that an alternative approach to this problem would be to collect additional data from an image search engine, and to turn to domain alignment techniques, such as the generate to adapt approach proposed by Sankaranarayanan et al. \cite{sankaranarayanan2018generate}, where a source dataset is aligning with a target domain using \textcolor{black}{Generative Adversarial Networks (GANs) \cite{goodfellow2014generative}}.}

\textcolor{black}{GANs are also increasingly being used as a technique for dataset augmentation~\cite{sixt2018rendergan}. This technique has relevant applications towards improving the classification of advertisements.} We note that advertisements extracted from street level imagery are often partially obscured by other real world entities (cars, trees, pedestrians).  
We propose to embed selected advertisements within street-level imagery through GANs to create additional training data (albeit `fake data') for model training. To date we can show that advertisements can be successfully integrated into street-level images. We place the advertisement using a STN to transform the image to a target shape. Finally we train GANs to realistically embed the images. 
We hypothesize that augmenting our collected street view data with these secondary GANs created data will enable the training of an effective model.



\section{Conclusion} \label{sec:discussion}

Our study presents a novel open deep learning workflow for extracting and classifying unhealthy advertisements within street-level imagery. Tackling inequalities in exposures to unhealthy advertisements might offer feasible regulatory opportunities for public authorities, especially when coupled with efficient and effective data collection methods to support decision making. There are very few to no existing secondary datasets providing this information to public authorities or researchers, and our project can solves this barrier to effective decision making. Our deployable tool can be used to efficiently collect data for understanding exposure to unhealthy advertisements, as well as identifying areas with high exposures that can benefit from restriction policies. 

\section{Acknowledgements}

This work was supported by the Economic and Social Research Council [grant number ES/L011840/1]. 

\section{Author Contributions}

G.P, M.G, and  A.S designed the research, analyzed the results and wrote the paper. E.B made key contributions with respect to the literature review. R.S made conceptual suggestions and also contributed towards the writing of the paper. Y.V made contributions towards the acquisition and preparation of data. All authors reviewed the manuscript.    

\section{Additional Information}

\noindent{\textbf{Supplementary information}} regarding accessing the data and deep learning frameworks discussed in our work can be found here: \href{https://github.com/gjp1203/LIV360SV}{https://github.com/gjp1203/LIV360SV}.

\noindent{\textbf{Competing Interests:}} The authors declare no competing interests.





\bibliography{bibliography.bib}







\end{document}